\documentclass[aps,pra,preprint,groupedaddress,showpacs,showkeys]{revtex4}
\usepackage{amsthm,amsfonts,amsmath,graphicx,subfigure}

\begin{document}

\title{ANOMALOUS DIFFUSION ON CRUMPLED WIRES IN TWO DIMENSIONS}
\author{C.C. Donato}
\author{F.A. Oliveira}
\affiliation{Instituto de F\'{\i}sica, and CIFMC, Universidade de
Bras\'{\i}lia, 70910-900, Bras\'{\i}lia, DF, Brazil.}
\author{M.A.F. Gomes\footnote{Corresponding author. Fax 55 81 3271 0359}}
\affiliation{Departamento de F\'{\i}sica, Universidade Federal de Pernambuco, 50670-901, Recife, PE, Brazil.}
\email{mafg@ufpe.br (M.A.F. Gomes)}

\begin{abstract}
It is investigated the statistical properties of random walks evolving on real configurations of a crumpled wire rigidly jammed in two dimensions. These crumpled hierarchical structures with complex topology are obtained from a metallic wire injected at a constant rate into a transparent planar cell of $20cm$ of diameter. The observed diffusion is anomalous with an exponent very close to that obtained at the threshold of two dimensional percolation. A comparison of the system studied in this paper with other systems of physical interest is also made, and an experimental consequence of our results is discussed.
\end{abstract}


\maketitle

\section{\label{sec1}Introduction}

The study of crumpled structures is a topic of increasing scientific interest in physics~\cite{ref1,ref2} and correlated disciplines~\cite{ref3}, as well as in the mathematical and technological domains~\cite{ref4}. In the last years, theoretical and experimental aspects of the condensed matter physics of crumpled sheets have been investigated in many areas of study, e.g. acoustic emission~\cite{ref5}, continuous mechanics~\cite{ref6}, growth models~\cite{ref7}, packing problems~\cite{ref8}, polymer, membrane, and interface physics~\cite{ref9}, universality~\cite{ref10}, among others. However, crumpled structures with different topologies, as exemplified by a squeezed ball of wire, have been much less studied in the physics literature. Geometrical, statistical, and physical aspects of crumpled wires ({\em CW}) in $3d$ space were examined fifteen years ago from the point of view of experimental work and analogue simulations, and in particular some robust scaling laws and fractal dimensions associated with these disordered systems were reported~\cite{ref11}. More recently, it was discovered that the irreversible process of injection of a long metallic wire in a two dimensional cavity lead to crumpled structures exhibiting several scaling phenomena~\cite{ref12}. In the present work, we report results of an extensive analysis of diffusion on these $2d$ crumpled structures of wire obtained by irreversible squeezing of macroscopic pieces of copper wires within a two-dimensional transparent cell.

The outline of this article is as follows. In Sec.~\ref{sec2} we describe the essential details from the experiment and from the simulations, and in Sec.~\ref{sec3} we present and discuss our results. In Sec.~\ref{sec4} we summarize our major conclusions.

\section{\label{sec2}Experimental details and simulations}

\begin{figure}
 \resizebox{8cm}{!}{\includegraphics{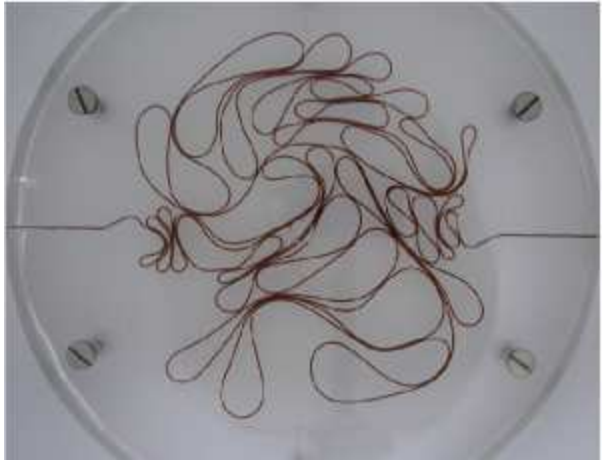}}
 \caption{\label{fig1} Photograph of the $2d$ injection cell used in the experiments discussed in this paper. The internal cavity contains a typical {\em CW} configuration (in this particular case for the maximum packing fraction, with $14.0\%$ of the area of the cavity occupied by the {\em CW}). All configurations of {\em CW} in $2d$ present a hierarchical cascade of loops with several types of contact wire-wire leading to a complex topology. See Sec.~\ref{sec2} for detail.}
\end{figure}

The experimental apparatus used in our work to obtain the configurations of $2d$ {\em CW} consists of a transparent cell formed by the superposition of two discs of plexiglass with a total height of $1.8cm$, an external diameter of $30cm$ and an internal circular cavity with $R = 10.0cm$ of radius and $0.11cm$ of height, which can accommodate configurations of a {\em single} layer of {\em CW} of $0.10cm$ of diameter. The cavity of the cell was polished and the copper wire used in the experiments (\#19AWG, with diameter $\zeta = 0.10cm$) had a varnished surface, in order to reduce the friction. Cavity and wire operated in dry regime, free of any lubricant. Two radial channels were made to provide the injection of wire into the cell at the angle of $180^{\circ}$ , as indicated in Fig.\ref{fig1}. Digital photographs of the {\em CW} configurations were taken with an Olympus C-3040ZOOM digital camera with resolution of 3 megapixels, which was connected with a PC, and assembled $30cm$ over the cell. To avoid picture artifacts by light reflections a cylindrical paper screen was placed around the cell.

In order for the reader to develop insight about the nature of the crumpling processes that we are considering in this paper, we show in Fig.\ref{fig1} a typical $2d$ configuration of {\em CW} at the limit of maximum packing corresponding to an occupation of $14.0\%$ of the total area of the cell. The packing fraction is given $p =  \frac{\zeta L}{\pi R^2}$, where $L$ is the length of the wire introduced in the cavity. The maximum length of wire which is possible to introduce in this case is $L_{max} = 440cm$ (ensemble average), corresponding to $p_{max}  = 0.14$. In fact, this maximum packing fraction does not vary appreciably, because it is controlled by the (approximately universal) ratio $\left(\frac{\mbox{shear modulus}}{\mbox{modulus of elasticity}}\right)$ of the wire~\cite{ref13}. The maximum packing fraction is also independent of the degree of lubrication of the cavity~\cite{ref12}. Each injection experiment begins fitting a straight wire in the opposite channels and subsequently pushing manually and uniformly the wire on both sides of the cell toward the interior of the cavity. The injection velocity at each channel in these experiments was of the order of $1cm/s$. For occupations close to $p_{max}$, the crumpled structures become rigid, the difficulty in the injection increases, and the injection velocity goes to zero abruptly. However, the observed phenomena are widely independent of the injection speed for all interval of injection velocity compatible with a manual process. The complex crumpled patterns of wire observed within the cavity are basically due to the formation of a cascade of loops of decreasing size. During the progressive injection of wire into the cell, the cascade of loops evolves in such a way that it is common to observe localized or global rearrangements of the loops previously formed, provided the packing fraction is not close to $p_{max}$. In principle, there is a concentration of stress and energy in the regions of the loops with smaller curvature radii, and a negligible amount of stretching is involved. The reader can also observe that the sharp creases and ridges found in crumpling of sheets are absent in the $2d$ {\em CW} shown in Fig.~\ref{fig1}. For additional details on the preparation and geometric properties of $2d$ {\em CW} we refer the interested reader to reference~\cite{ref12}.

\begin{figure}
 \resizebox{10cm}{!}{\includegraphics{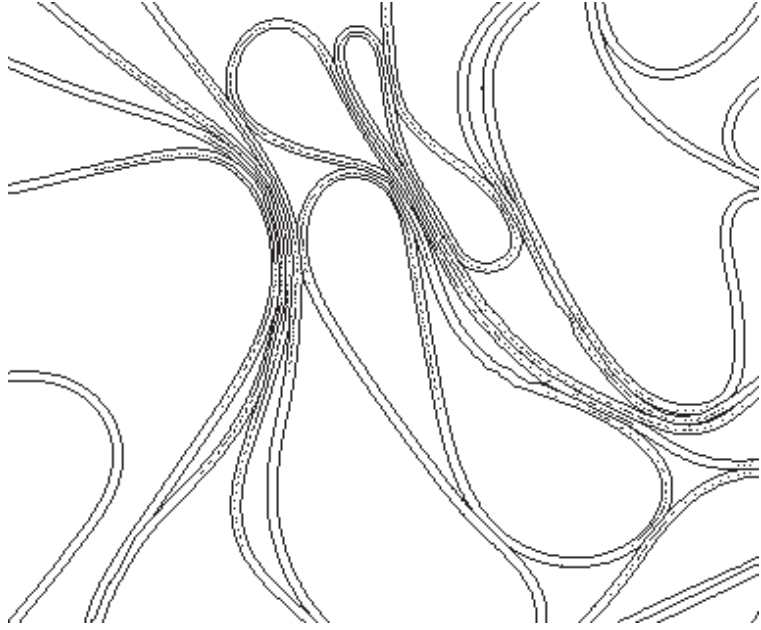}}
 \caption{\label{fig2} A detail of Fig.~\ref{fig1} showing the geometry of the contacts wire-wire. The distribution of points within the trail (wire backbone) represents the positions of a random walk of 10,000 steps. It can be noticed that the walk is trapped in a very restricted region of the {\em CW} - an indication that the process is diffusive, with an anomalous diffusion exponent $d_w > 2$. See Secs.~\ref{sec2} and~\ref{sec3} for detail.}
\end{figure}

Using high resolution digital images from an ensemble of five independent and equivalent configurations of {\em CW} at the limit of maximum packing (Fig.~\ref{fig1}) we simulate random walks by considering the following rules: (i) The random walker has size $\zeta$  identical to the width of the wire. (ii) Each walk begins in a point chosen at random on the {\em CW}. (iii) In regions free of contact wire-wire, the random walker at position $r_i$ had equal probabilities to go ahead (to $r_{i+1}$) or to go back (to $r_{i-1}$) in unit steps, as in a typical one-dimensional trail. (iv) On the other hand, in regions with contact wire-wire, the more complex local topology must be considered and the walker can move with equal probability from $r_i$ to one of its closest neighbors sites $\{\delta\}$ (i.e. a number of neighbors~$ > 2$) localized at $r_{i+\delta}$. For illustration, we show in Fig.~\ref{fig2} a detail of a region taken from Fig.~\ref{fig1} in which a large number of contacts wire-wire occur. In these regions the walker can jump from a loop to another contiguous loop, and this possibility modifies drastically the topology of the random walk. From Fig.~\ref{fig2} one can notice contacts of variable extent involving e.g. 2, 3, 4, and 5 trails.

\section{\label{sec3}Results and discussion}

\begin{figure}[t!]
 \resizebox{13cm}{!}{\includegraphics{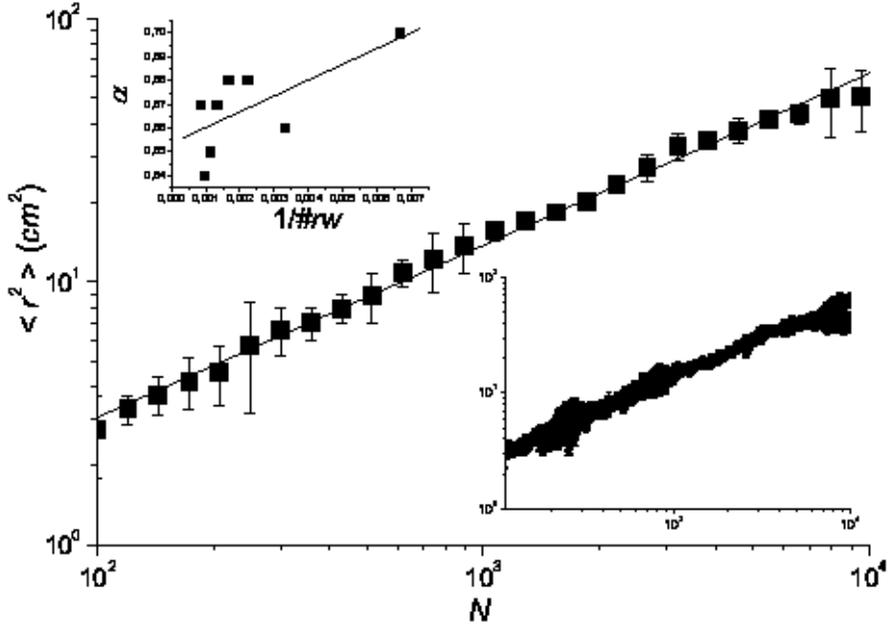}}
 \caption{\label{fig3} The main plot shows the $log-log$ dependence of the mean square displacement for random walks on a {\em CW},  $<r^2(N)>$, as a function of the number of steps $N$ (only 26 selected points are represented to put in evidence the power law fit and the fluctuation bars. The data follows the scaling  $<r^2(N)> \sim N^\alpha$, with $\alpha$   assuming the value $0.66 \pm 0.01$ (obtained from the interval $100 < N < 3,000$). The random walk exponent $d_w^{(CW)} \equiv \frac{2}{\alpha}$  assumes the anomalous value $3.03 \pm 0.05$. The upper inset shows the dependence of the exponent $\alpha$ with the number of random walks used in the ensemble. The lower inset shows the same variables of the main plot, but for the complete simulation data. See Sec.~\ref{sec3} for detail.}
\end{figure}

\begin{figure}[t!]
 \centering
\subfigure[\label{fig4a}]{\resizebox{11cm}{!}{\includegraphics{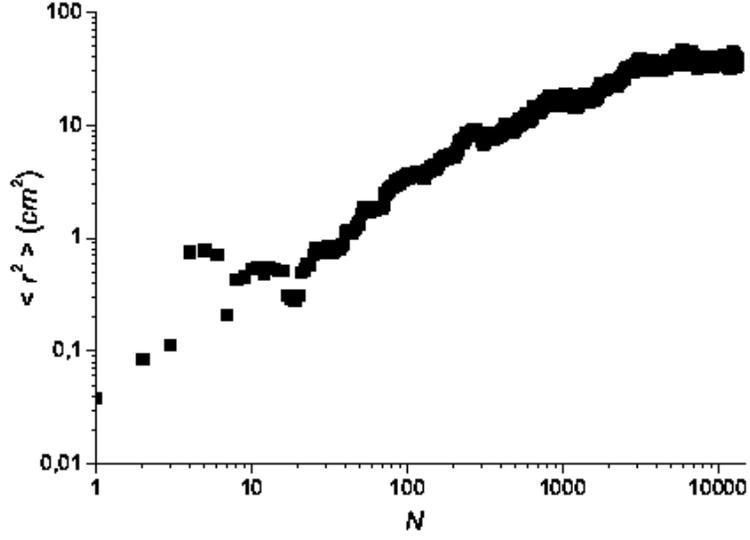}}}
\subfigure[\label{fig4b}]{\resizebox{11cm}{!}{\includegraphics{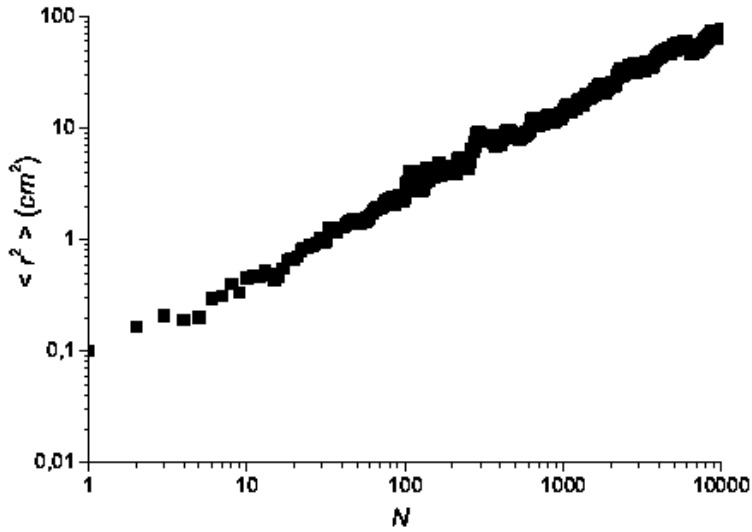}}}
 \caption{\label{fig4} (a) and (b) show the mean square displacement of a random walker for two individual samples (see text, Section~\ref{sec3}, for detail).}
\end{figure}

Due to the possibility of diffusion transverse to the trails in the regions with a variable number of contacts between loops, an extensive study of the statistical properties of the random walks on {\em CW} is necessary. Using the rules defined in the end of Sec.~\ref{sec2}, we calculated the mean square displacement of a random walker,  $<r^2(N)>$, as a function of the number of steps $N$ as shown in the $log-log$ plot of Fig.~\ref{fig3}. This ensemble average was obtained from 1,150 walks of 10,000 steps for all configurations of {\em CW} considered. We see from the main plot of Fig.~\ref{fig3} that $<r^2(N)>$ is well described by a power law function: $<r^2(N)> \sim N^\alpha$, with the slope $\alpha$  assuming the value $0.66 \pm 0.01$ (Obtained from the interval $100 < N < 3,000$; i.e. for relatively not too large number of steps ["short-time" limit] to avoid the finite size effect of the cavity associated with the longer walks.). Thus, the random walk exponent $d_w^{(CW)} \equiv \frac{2}{\alpha}$  assumes the anomalous value $3.03 \pm 0.05$, sensibly larger than the brownian exponent $d_w^{(B)} = 2$ for normal diffusion. In order to display clearly the power law best fit leading to the determination of the exponent $\alpha$ (continuous line) and the typical fluctuation bars in the main plot, we exhibit in this figure only 26 data points. The complete overcrowded simulation data is represented in the lower right inset. For the sake of completeness we show in the upper left inset of Fig.~\ref{fig3} the dependence of the exponent  $\alpha$ with the inverse of the number of random walks ({$\#rw$}) used in the simulations, for some values of $\#rw$ in the interval [150, 1150]. The linear fit (continuous line) in this inset suggests that $\alpha$  converges to the value 0.655 within uncertainties of 0.010 at the limit $\#rw \rightarrow \infty$. Obviously, a study of the finite-size scaling for the random walks along the line suggested in reference~\cite{ref14} would be interesting in this problem. However, the realization of experiments with cavities of several different sizes is presently connected with severe technical difficulties. Figure~\ref{fig4a} and~\ref{fig4b} give for completeness additional information on the sample to sample fluctuations on $<r^2(N)>$: in Fig.~\ref{fig4a}, differently from Fig.~\ref{fig4b}, the saturation on $<r^2(N)>$ is already evident for $N$ close to $10,000-13,000$ steps. We notice that the exponents $d_w$ obtained from the single {\em CW} configurations exhibited in these figures are close to the exponent found from Fig.~\ref{fig3}. In particular, from Fig.~\ref{fig4b} we obtain $d_w = 2.86 \pm 0.04$, for $10 < N < 10^4$.

The value obtained from Fig.~\ref{fig3} for the exponent $d_w^{(CW)}$ is close to the value $d_w^{(PC)}~=~2.87~\pm~0.02$ obtained for anomalous diffusion on the self-similar incipient infinite percolation cluster ({\em PC}) at $d=2$~\cite{ref15}. Although $2d$ {\em CW} and $2d$ percolation clusters have different geometric aspects, both structures possess some similarities: {\em CW} has fractal dimension $D^{(CW)} = 1.9 \pm 0.1$~\cite{ref12} (The reader can observe that this is the {\em experimental} value obtained from the {\em CW} configurations studied in~\cite{ref12}; i.e. exactly the {\em same} {\em CW} configurations used in the present work.), and {\em PC} has a close fractal dimension $D^{(PC)} = 1.89 \pm 0.03$; they have structures with the same topological dimension, $d_T = 1$; they have a hierarchy of holes or empty spaces without any characteristic size, and they equably can support stress.

The exponent $d_w^{(CW)}$ reported here for {\em CW} is in agreement with the conjecture of Aharony and Stauffer~\cite{ref16} which predicts that $d_w = D + 1$, for a system with fractal dimension $D < 2$. According to this conjecture, the observed experimental value $D^{(CW)} = 1.9 \pm 0.1$~\cite{ref12}, points to a value $d_w^{(CW)} = 2.9 \pm 0.1$, which is equal to $d_w^{(CW)} = 3.03 \pm 0.05$ within the statistical fluctuations.

As an additional insight, it is interesting to compare the exponent $d_w^{(CW)}$ with the diffusion exponent for a $2d$ self-avoiding random walk ({\em SARW}) with a (random) distribution of local bridges without mass connecting any two nearest-neighbor sites visited by the {\em SARW}. This structure which could be speculated in principle as a valid zero-order approximation for $2d$ {\em CW} allows both diffusion along the backbone of the {\em SARW} as well as diffusion transverse to the {\em SARW} backbone (via jumps of the walker along the bridges) was studied by Manna and Roy~\cite{ref17}. This {\em SARW} with bridges has the same fractal dimension of the {\em SARW}, that is $D = 1.33$. Manna and Roy found that $d_w = 2.56 \pm 0.01$ for the {\em SARW} with bridges, a value somewhat different albeit not too distant from $3.03 \pm 0.05$ observed for $2d$ {\em CW}.

An experimental consequence of the diffusion exponent reported in this paper is connected with the dependence of the electrical resistance $\Re$ of a {\em CW} with its size (radius) $R$. Here we are assuming a {\em CW} allowing for conduction along the wire backbone as well as transversal to it (in the regions of contact wire-wire). In general, the electrical resistance of an Euclidean $d$-dimensional system of size $R$ scales as $\Re \sim \frac{R}{Area} = \frac{R}{R^{(d-1)}} = R^{(2-d)}$. For a fractal system, the corresponding relation is $\Re \sim R^{(d_w-D)}$, where $D$ is the fractal dimension $D \leq d$, and $d_w$ is the random walk exponent obtained from random walks defined on the fractal~\cite{ref18}. The low connectivity of fractals lead to $d_w > 2$, i.e. to an exponent larger than the normal (brownian) value. Using the observed values $D^{(CW)}$ and $d_w^{(CW)}$ for {\em CW}, we predict that the electrical resistance for such systems scales as $\Re^{(CW)} \sim R^{(1.1 \pm 0.1)}$, whereas normal diffusion for a $2d$ Euclidean system implies that $\Re \sim R^{(2-2)} = R^0$. Thus, the electrical resistance of a {\em CW} of the type considered above behaves very close to the electrical resistance of a $1d$ resistor ($\Re \sim$~size). For comparison, the expected dependence electrical resistance versus size for a percolation cluster is $\Re \sim R^{0.98 \pm 0.05}$, obtained from the corresponding values of $d_w^{(PC)}$ and $D^{(PC)}$. On the other hand, the electrical resistance for a {\em CW} allowing for conduction only along the backbone of wire (i.e. excluding jumps of current in the contacts wire-wire) scales as $\Re \sim$~(length of wire within cavity)~$\sim R^D \sim R^{1.9 \pm 0.1}$.

Recently~\cite{ref19}, an investigation in the fluctuation of the density of states in a quantum Heisenberg disordered chain did show its equivalence with the noise spectral density within the calculation precision. This result strength the conjecture~\cite{ref20, ref21} which suggest the equivalence between both densities. Moreover, since we can predict the diffusion exponents from the noise~\cite{ref20, ref22}, it would be important to use this conjecture in {\em CW} systems to obtain those exponents in an independent way.

\section{\label{sec4}Conclusions}

Using extensive numerical simulations we have studied the statistical behavior of a random walker evolving directly on the experimental complex configurations of crumpled wires confined in a two dimensional cavity, at the limit of maximum packing fraction~(Fig.~\ref{fig1}). In particular, we have obtained the dependence of the averaged square distance  $<r^2(N)>$ of the random walks as a function of the number of steps $N$. It is found that these quantities scale as  $< r^2(N) > \sim N^\alpha$, where $\alpha = \frac{2}{d_w}  = 0.66 \pm 0.01$, corresponding to a diffusion exponent $d_w^{(CW)} = 3.03 \pm 0.05$, when the packing fraction of wire in the cavity approaches the limit value $0.14 \pm 0.01$. This value for the diffusion exponent is close to that obtained for diffusion on the two dimensional incipient percolation cluster~\cite{ref15}. The exponent $d_w^{(CW)}$ implies that the electrical resistance of crumpled wires scales with the radius (size) as {\bf $\Re^{(CW)} \sim R^{(1.1 \pm 0.1)}$}, a result that can be tested with precise electrical measurements.

\begin{acknowledgments}
This work was supported in part by Conselho Nacional de Desenvolvimento Cient\'{\i}fico e Tecnol\'ogico, Fundo Setorial do Petr\'oleo, Programa de N\'ucleos de Excel\^encia, and Coordena\c{c}\~ao de Aperfei\c{c}oamento de Pessoal de N\'{\i}vel Superior (Brazilian Agencies).
\end{acknowledgments}

\end{document}